\documentclass{PoS}
\DeclareMathAlphabet{\mathit}{OT1}{cmr}{m}{it}
\usepackage{wrapfig}
\usepackage{citesort}
\usepackage{subfigure}
\usepackage{upgreek}

\def\meannch{{\langle n_{\rm ch} \rangle}}

\title{Particle spectra at ZEUS}

\ShortTitle{Particle spectra at ZEUS}

\author{\speaker{Lydia Shcheglova} \thanks{for the ZEUS Collaboration}\\
        Nuclear Physics Institute of Moscow State University\\
        E-mail: \email{lydia.shcheglova@desy.de}}

\abstract{Charged multiplicity distributions and the mean charged
multiplicity  have been investigated in inclusive neutral current deep
inelastic $\mathrm{e}\mathrm{p}$  scattering with the ZEUS detector at
HERA using an integrated luminosity of $38.6\:\mathrm{pb}^{-1}$. The
measurements were performed in the current region of the Breit frame,
as well as in the current fragmentation region of the  hadronic
center-of-mass frame. The KNO-scaling properties of the data were
investigated and the energy dependence of multiplicity distributions
was studied using different energy scales. The data are compared to
results obtained in $\mathrm{e}^+\mathrm{e}^-$ collisions and  to
previous DIS measurements as well as to leading-logarithm
parton-shower Monte Carlo predictions. The scaled momentum
distributions of charged  particles in jets have been also measured
for dijet photoproduction with the  ZEUS detector at HERA using an
integrated luminosity of $359\:\mathrm{pb}^{-1}$.  The distributions
are compared to predictions based on perturbative QCD  carried out in
the framework of the modified leading-logarithmic  approximation
(MLLA) and assuming local parton-hadron duality (LPHD). The universal
MLLA scale, ${\mathit{\Lambda}}_{\mathrm{eff}}$, and the LPHD
parameter, $\kappa^{\mathrm{ch}}$, are extracted.}

\FullConference{European Physical Society Europhysics Conference on
High Energy Physics\\
                 July 16-22, 2009\\
                 Krakow, Poland}

\begin{document}

\section{Introduction}

The HERA ep collider provides a rich field for the study of particle production
in a wide range of $\mathit{W}$, the photon-proton centre-of-mass (CMS) energy,
and the photon  virtuality $Q^2$. The data presented here were obtained
with the ZEUS detector at $\sqrt{\mathit{s}} \sim 300\:{\mathrm{GeV}}^2$
and concern the study of the hadronisation and parton fragmentation
processes, phenomena which give deep insight into the non-perturbative
sector of QCD. Establishing universal features in the properties of
the final hadronic system in reactions with different initial particles
($\mathrm{e}^+\mathrm{e}^-$, $\mathrm{ep}$, hadron scattering) helps
to elucidate how the partonic cascades evolve into observed
hadrons. In particular, recent results on multiplicity distributions
in DIS as functions of different energy scales and detailed comparisons with
$\mathrm{e}^+\mathrm{e}^-$ are discussed below. The formation of hadron
jets was also investigated using the multiplicity and momentum spectra
of charged hadrons in the dijet photoproduction events. The measurements
verify the  validity and consistency of the MLLA approach at energy scale 
accessed at HERA.

The multipurpose ZEUS detector is described in detail elsewhere~\cite{1} 

\section{Multiplicity of charged hadrons}

The average multiplicity and multiplicity distributions are being
studied intensively in particle collisions. In  previous studies of
DIS events only the virtuality  of the exchanged photon, $Q$, was used
as the energy scale~\cite{2}. A reasonable agreement with
$\mathrm{e}^+\mathrm{e}^-$ data was shown except for the region of $Q$
below  $6-8\:\mathrm{GeV}$. Recently, the ZEUS collaboration has
performed a detailed study of the charged multiplicity in the neutral
current deep  inelastic scattering (DIS)~\cite{3}.   These
measurements of the charged hadron multiplicity are performed in the
Breit and in the hadronic centre-of-mass (HCM) frames. Due to the
restricted  detector acceptance only hadrons belonging to the current
fragmentation regions in both  frames were used in the analysis.
\begin{figure}[htb]
  \vspace{-10pt}
  \begin{minipage}[b]{0.48\linewidth}
    \begin{center}
      \includegraphics*[viewport=25 10 530 560, width=0.98\textwidth]{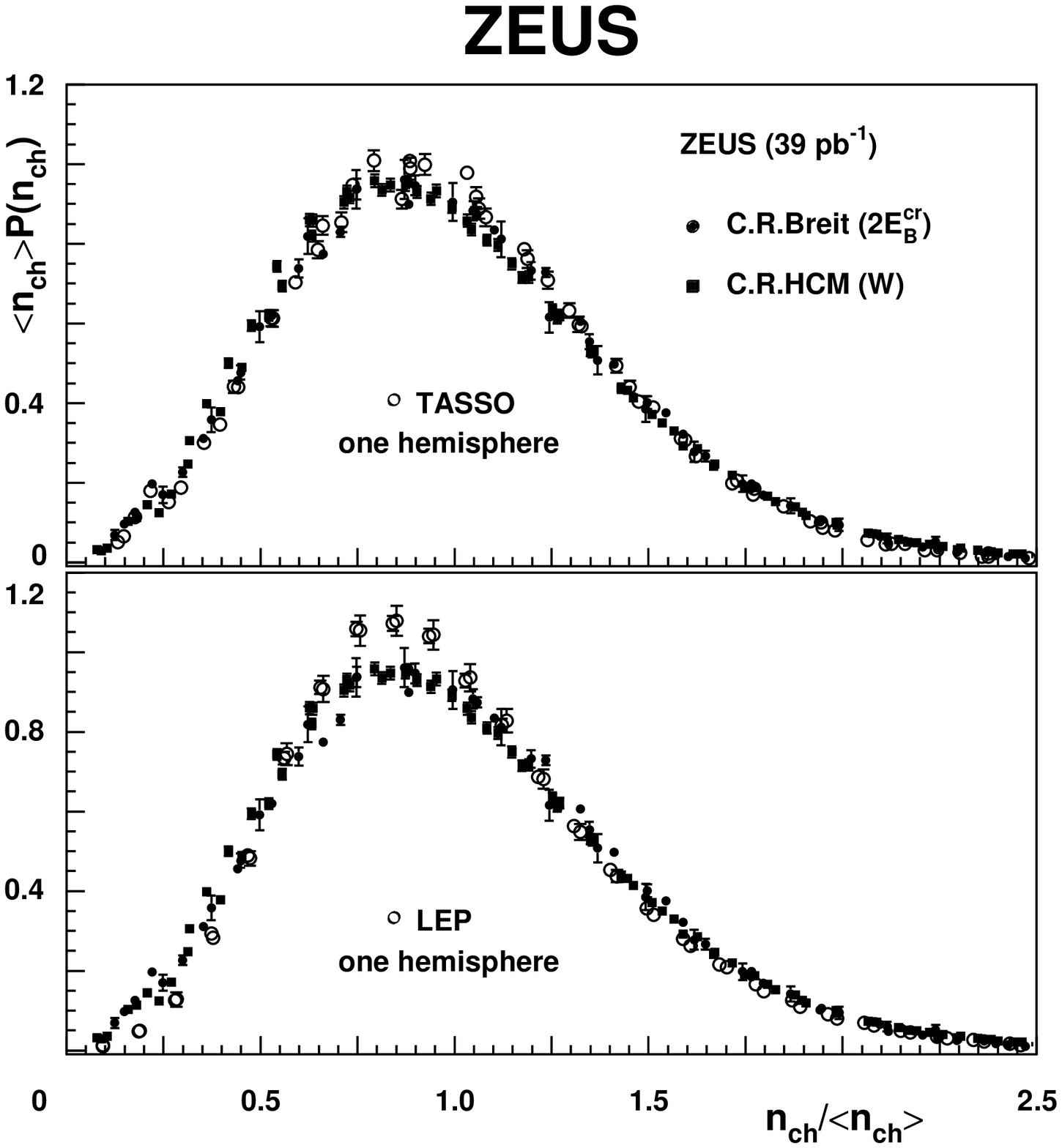}
    \end{center}
    \vspace{-20pt}
    \caption{Comparison of multiplicity distributions in KNO form
             in bins of $2\cdot E^{\mathrm{cr}}_{\mathrm{B}}$ and 
             $W$ (solid markers) with $\mathrm{e}^+\mathrm{e}^-$ data for one hemisphere
             (open circles).}
    \label{fig:kno_01}
  \end{minipage}
  \hfill
  \vspace{-10pt}
  \begin{minipage}[b]{0.48\linewidth}
    \begin{center}
      \includegraphics*[viewport=25 10 530 560, width=0.98\textwidth]{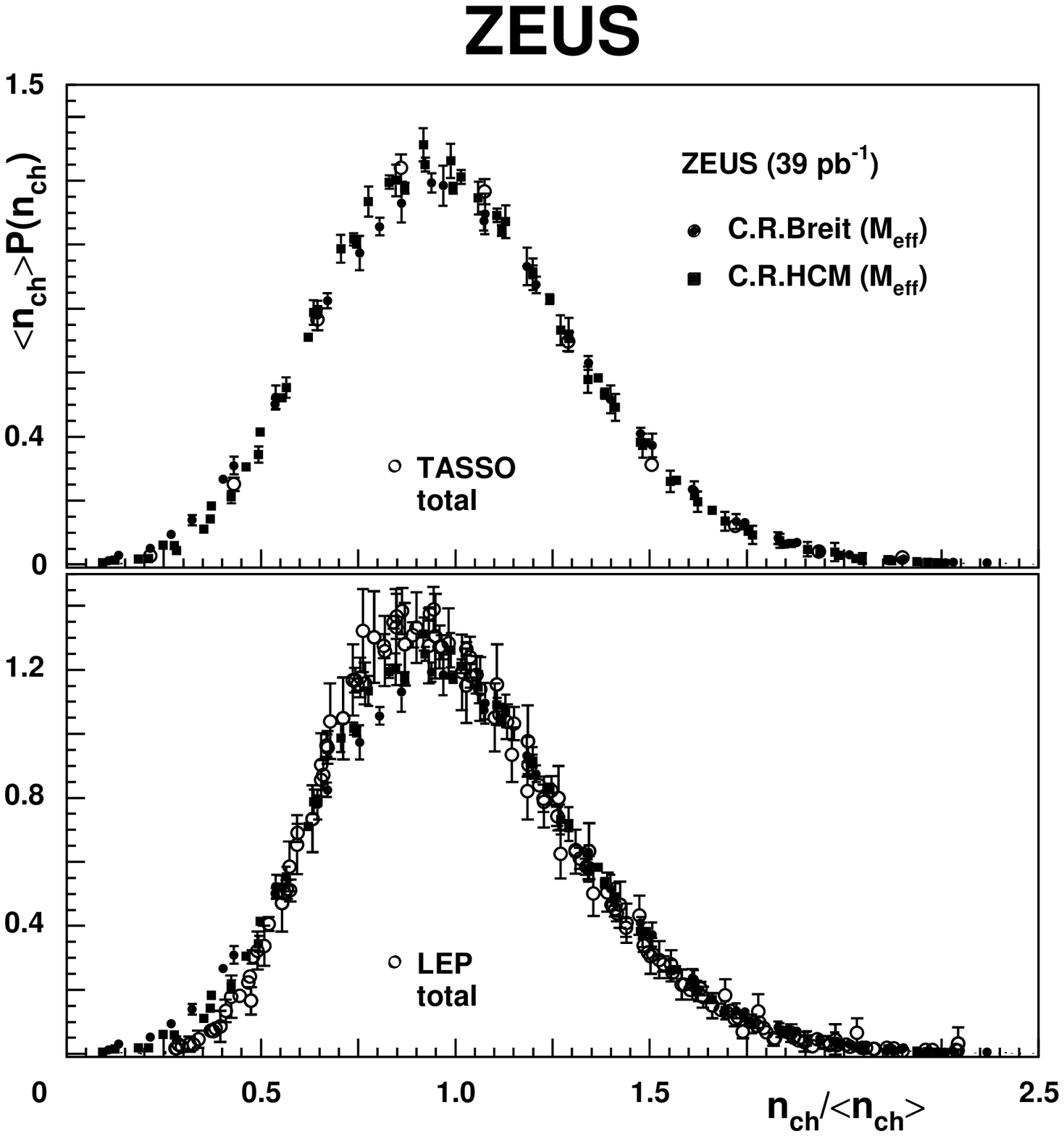}
    \end{center}
    \vspace{-20pt}
    \caption{Comparison of multiplicity distributions in KNO
    form in bins of $M_{\mathrm{eff}}$ (solid markers)
     with $\mathrm{e}^+\mathrm{e}^-$ data for the whole event
    (open circles).} \label{fig:kno_02}
  \end{minipage}
\end{figure}

\begin{wrapfigure}{r}{0.5\textwidth}
  \vspace{-30pt}
  \begin{center}
    \includegraphics*[viewport=25 10 530 560, width=0.48\textwidth]{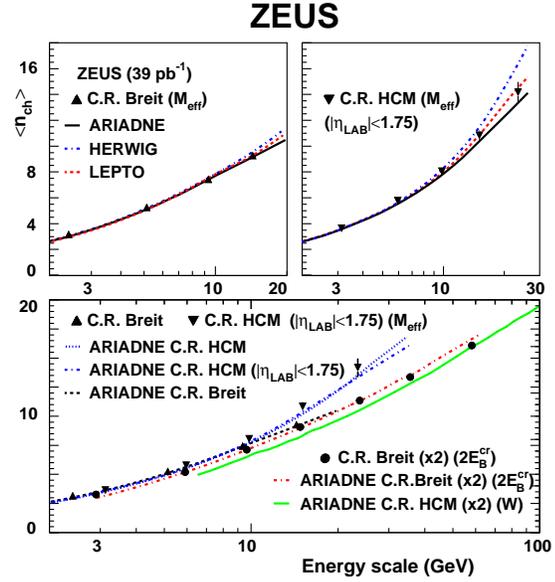}
  \end{center}
  \vspace{-20pt}
  \caption{Mean charged multiplicities as a function of $M_{\mathrm{eff}}$
           compared to MC predictions (a, b). Comparison of the measurements
           as functions of $M_{\mathrm{eff}}$ with the measurements
           as a function of $2\cdot E^{\mathrm{cr}}_{\mathrm{B}}$ along with the MC
           predictions (c).}
\label{fig:energy_scale_1}
\vspace{-10pt}
\end{wrapfigure}
ZEUS investigated charged multiplicity distributions and mean charged
multiplicity in terms of different energy scales
in order to consistently compare  $\mathrm{e}\mathrm{p}$ DIS data 
with the data from  $\mathrm{e}^+\mathrm{e}^-$, $\upnu\mathrm{p}$ and
$\upmu\mathrm{p}$ scattering. The following alternatives to the $Q$ energy
scale were considered: the energy of the current region  of the Breit
frame $E^{\mathrm{cr}}_{\mathrm{B}}$, the invariant mass of the produced
particles  $W$, used in the current region of HCM, and the invariant
mass of the hadronic system $M_{\mathrm{eff}}$, used in both frames.

The scaling properties of multiplicity distributions in a commonly
used form, proposed by Koba-Nielsen-Olsen (KNO)~\cite{4}, were studied
in bins of $W$, $2\cdot E^{\mathrm{cr}}_{\mathrm{B}}$ and
$M_{\mathrm{eff}}$ and  compared with $\mathrm{e}^+\mathrm{e}^-$ data.
The multiplicity distributions  in the KNO form from ZEUS are shown in
Figs.~\ref{fig:kno_01},~\ref{fig:kno_02}. In these plots the scaled
multiplicity distributions, $\Psi(z) = \meannch P(n_{\rm ch})$, are
plotted as a function of $n_{\rm ch} / \meannch $, where $P(n_{\rm
ch})$ and $\meannch$ are the multiplicity distribution and average
multiplicity respectively. Fig.~\ref{fig:kno_01} shows a comparison of
the KNO distributions in bins of $2\cdot E^{\mathrm{cr}}_{\mathrm{B}}$
($12 < 2\cdot E^{\mathrm{cr}}_{\mathrm{B}} < 100\:\mathrm{GeV}$) and
in bins of $W$ ($ 70 < W < 225\:\mathrm{GeV}$) with measurements in
one hemisphere  of $\mathrm{e}^+\mathrm{e}^-$, obtained by the TASSO
collaboration in the energy range  $14 < \sqrt{s_{\mathrm{ee}}} <
44$~\cite{5} and by the LEP experiments at $\sqrt{s_{\mathrm{ee}}} =
91.2\:\mathrm{GeV}$~\cite{b6,b7}. There is a  remarkable agreement
between  $\mathrm{ep}$ and TASSO data; the LEP data differ somewhat
from the present measurement in the peak region and at very low 
$n_{\rm ch} / \meannch$.

\begin{wrapfigure}{l}{0.5\textwidth}
  \vspace{-25pt}
  \begin{center}
    \includegraphics*[viewport=15 15 515 550, width=0.48\textwidth]{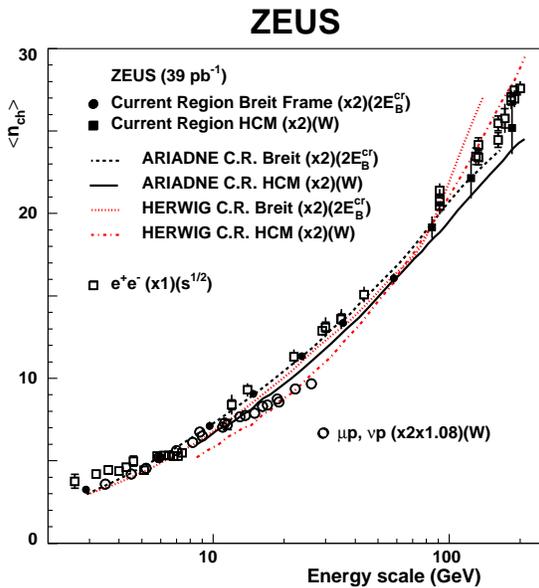}
  \end{center}
  \vspace{-25pt}
  \caption{ Comparison of mean charged multiplicities as a function of
           different energy scales with the data from $\mathrm{e}^+\mathrm{e}^-$
           and fixed target experiments.}
  \label{fig:energy_scale_2}
  \vspace{-20pt}
\end{wrapfigure}
A comparison of the KNO distributions with
$\mathrm{e}^+\mathrm{e}^-$ data (both hemispheres) in
$M_{\mathrm{eff}}$ bins ($8 < M_{\mathrm{eff}} < 30\:\mathrm{GeV}$)
in  the current region of HCM and in the current region of the Breit
frame ($8<M_{\mathrm{eff}}< 20\:\mathrm{GeV}$) is shown in
Fig.~\ref{fig:kno_02}. There is good agreement between the ZEUS  and
both TASSO~\cite{5} and LEP data~($91.2 < \sqrt{s_{\mathrm{ee}}} <
209\:\mathrm{GeV}$)~\cite{b6,b7,b8}.
The mean charged multiplicities
were also investigated using the different energy scales discussed above.
In Fig.~\ref{fig:energy_scale_1} the mean charged multiplicities,
measured in the current regions of the Breit and HCM frames, are
presented as functions of the invariant mass of the corresponding
hadronic system and compared with the MC predictions.
All three MC models describe the data reasonably well although the Herwig
prediction is  too high in the last bin of $M_{\mathrm{eff}}$ in the
current region of the HCM  (Figs.~\ref{fig:energy_scale_1}~(a),~(b)). As
is seen in Fig.~\ref{fig:energy_scale_1}~(c), the data in the Breit and
HCM frames agree in the region of $M_{\mathrm{eff}} < 10\:\mathrm{GeV}$,
while at higher $M_{\mathrm{eff}}$, $\meannch$ rises much faster with
$M_{\mathrm{eff}}$ in the current region of HCM than in the current
region of the Breit frame. In Fig.~\ref{fig:energy_scale_1}~(c)  the
values of  $2\cdot {\meannch}$ as a function of $2\cdot
E^{\mathrm{cr}}_{\mathrm{B}}$ are also plotted. The data  follow the
same dependence as $\meannch$ vs. $M_{\mathrm{eff}}$ in the Breit frame
but differ  from those obtained in the current region of the HCM.

Finally,  Fig.~\ref{fig:energy_scale_2} shows the comparison of the the
mean  charged multiplicities in the  current region of the Breit and HCM
frames  as a function of $2\cdot E^{\mathrm{cr}}_{\mathrm{B}}$ and
$W$ with the data from $\mathrm{e}^+\mathrm{e}^-$~ and fixed-target 
experiments. The fixed-target data were scaled by a factor 2 (since they
were measured in one hemisphere only) and were corrected for the
${\mathrm{K}}^0_{\mathrm{S}}$ and $\Lambda$ decays by a factor 1.08,
estimated using the ARIADNE MC model.The ZEUS measurements show good overall
agreement with the data from other experiments and exhibit approximately
the same dependence on the respective energy scale; only the fixed
target DIS data deviate at energies above $15\:\mathrm{GeV}$. The
energy scale $2\cdot E^{\mathrm{cr}}_{\mathrm{B}}$ gives better
agreement with $\mathrm{e}^+\mathrm{e}^-$ data at low values of energy
than $Q$. The measurements of $\meannch$ as a function of $W$ agree, within 
the uncertainties, with the data from $\mathrm{e}^+\mathrm{e}^-$ collisions.

\section{Scaled momentum distributions}

Recently, momentum spectra of charged hadrons in photoproduction were
studied in jet fragmentation processes with the   ZEUS
detector~\cite{b9}. The results are compared with perturbative QCD
calculations carried out in the framework of the Modified Leading Log
Approximation, MLLA,~\cite{b10} and the hypothesis of Local
Parton-Hadron Duality, LPHD~\cite{b11}.The MLLA equations give an
analytical description of the parton shower evolution and an effective
scale parameter of the QCD calculations,
${\mathit{\Lambda}}_{\mathrm{eff}}$, that is assumed to be universal,
i.e.\@ independent of the process considered. The LPHD hypothesis
predicts that the observed hadron distributions should be related to
the calculated parton distributions by a normalisation parameter
$\kappa^{\mathrm{ch}}$. Tests of the MLLA predictions in conjunction
with the LPHD hypothesis permit to expand our understanding of the
underlying physics of jet fragmentation phenomenon.

\begin{wrapfigure}{l}{0.5\textwidth}
  \vspace{-20pt}
  \begin{center}
    \includegraphics*[width=0.48\textwidth]{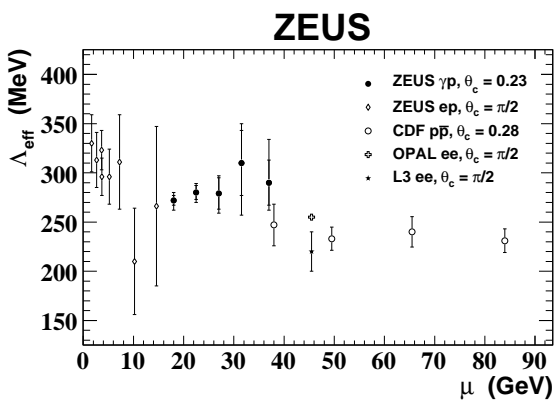}
  \end{center}
  \vspace{-20pt}
  \caption{${\mathit{\Lambda}}_{\mathrm{eff}}$ as a function of $\mu$,
           where $\mu$ denotes the characteristic energy scale for each
           specific process.}
  \label{fig:scalmom-01}
  \vspace{-10pt}
\end{wrapfigure}
Related studies~\cite{b12} have been performed before in
$\mathrm{e}^+\mathrm{e}^-$ collisions at LEP and PETRA, in DIS
$\mathrm{ep}$ collisions at HERA, (anti)neutrino-nucleon interactions
from the NOMAD experiment and $\mathrm{p}\bar{\mathrm{p}}$ collisions
at the Tevatron. In the analysis presented here, the momentum spectra
of charged hadrons are studied in dijet photoproduction ($\upgamma
\mathrm{p}$) events from $\mathrm{ep}$ collisions. The scaled momentum
distributions $\xi=\ln\left(1/x_\mathrm{p}\right)$, where
$x_\mathrm{p}$ is the fraction of the jet's momentum carried by the
charged particle, were measured in restricted cones of various opening
angles $\theta_c$ around the jet axis. Jets were reconstructed from
energy-flow  objects~\cite{b13} (EFOs) by  applying the
$k_\mathrm{T}$ cluster algorithm~\cite{b14}. The reconstructed invariant dijet
mass was used as an energy scale. It probes the range 19 to
$38\:\mathrm{GeV}$, which spans the energy region between those
accessed previously by the ZEUS and CDF collaborations.

To check the validity of the MLLA predictions using the measured $\xi$
distributions, two  approaches were adopted. The first was based on the
position of the peak of the $\xi$  distributions, $\xi_{\rm peak}$. The
second was based on a fit of the full shape of the $\xi$ 
distributions; the limiting spectra, predicted by MLLA+LPHD
theory~\cite{b10}, were used in the fit in this method. 

In the $\xi_{\rm peak}$ analysis the values of $\xi_{\rm peak}$ were
extracted from the $\xi$ distributions using a three-parameter
Gaussian fit. At leading order (LO), the peak position is predicted to
be at\\ $\xi_{\rm peak}=\frac{1}{2}Y + \sqrt{cY}-c$, where $c=0.29$
and Y is a function of the jet energy $E_{\rm jet}$ and $\theta_c$
(see eq.(3) in~\cite{b9}) and depends also on the parameter
${\mathit{\Lambda}}_{\mathrm{eff}}$. Thus the peak position can be
directly fit to the data, treating ${\mathit{\Lambda}}_{\mathrm{eff}}$
as a free parameter. The best fit value was found to be
${\mathit{\Lambda}}_{\mathrm{eff}}=275
\pm4$(stat.)$^{+4}_{-8}$(syst.)~$\mathrm{MeV}$ for $\theta_c$=0.23. 
In Fig.~\ref{fig:scalmom-01} the values of
${\mathit{\Lambda}}_{\mathrm{eff}}$ are shown as a function of the
energy scale and compared to the results from different experiments.
The data are consistent with the prediction that
${\mathit{\Lambda}}_{\mathrm{eff}}$ is a universal parameter.

\begin{wrapfigure}{r}{0.5\textwidth}
  \vspace{-20pt}
  \begin{center}
    \includegraphics*[width=0.48\textwidth]{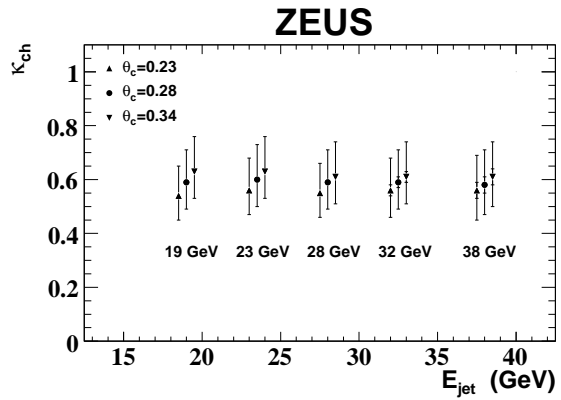}
  \end{center}
  \vspace{-20pt}
  \caption{$\kappa^{\mathrm{ch}}$ as a function of $E_{\rm jet}$ for the three
           $\theta_c$ values.
           }
  \label{fig:scalmom-02}
  \vspace{-10pt}
\end{wrapfigure}

The $\xi$  distributions were also fitted using the limiting spectrum
predicted by MLLA. The values of ${\mathit{\Lambda}}_{\mathrm{eff}}$
extracted from these MLLA fits are in reasonable agreement with those
extracted from the $\xi_{\rm peak}$ data, although the values obtained
using the MLLA fit have larger uncertainties due to sensitivity of
${\mathit{\Lambda}}_{\mathrm{eff}}$ to the choice of the fitting
range. The values of the LPHD parameters  $\kappa^{\mathrm{ch}}$ were
extracted also as a function of $E_{\rm jet}$ and $\theta_c$ from the
fitted limited momentum spectra and are shown in Fig.~6.  The value of
$\kappa^{\mathrm{ch}}$, measured with  $\theta_c=0.23$ and averaged
over $E_{\rm jet}$, was $\kappa_{\rm ch}=0.55
\pm0.01\mathrm{(stat.)}^{+0.03}_{-0.02}\mathrm{(syst.)}^{+0.11}_{-0.09}\mathrm{(theo.)}$
and is in good agreement with that reported by CDF collaboration,
$\kappa_{\rm ch}=0.56 \pm 0.05\mathrm{(stat.)}\pm
0.09\mathrm{(syst.)}$. The ZEUS data support the predicted
universality  of $\kappa_{\rm ch}$.

\section{Summary and conclusions}

The charged multiplicity distributions and the mean charged
multiplicity have been investigated in NC DIS $\mathrm{ep}$ scattering
in terms of different energy scales. Multiplicity distributions in the
scaling KNO form in the current regions  of the Breit and HCM frames
exhibit the same behaviour as those in one hemisphere of
$\mathrm{e}^+\mathrm{e}^-$ collisions when  $2\cdot
E^{\mathrm{cr}}_{\mathrm{B}}$ or $W$ are considered. When energy scale
$M_{\mathrm{eff}}$ is used, the charged  multiplicities exhibit the
same KNO-scaling behaviour as those for the whole
$\mathrm{e}^+\mathrm{e}^-$ event. The energy scales $2\cdot E^{\mathrm{cr}}_{\mathrm{B}}$   
and $W$ give better agreement with $\mathrm{e}^+\mathrm{e}^-$ data than $Q$.

The multiplicity distributions of charged particles in dijet
photoproduction events have been measured as a function of
$\xi=\ln\left(1/x_\mathrm{p}\right)$. Two methods, the $\xi_{\rm peak}$
analysis and fit of the $\xi$ distributions to the MLLA functions, were used to extract 
the MLLA scale,
${\mathit{\Lambda}}_{\mathrm{eff}}$, and LPHD parameter,
$\kappa^{\mathrm{ch}}$. The data support the assumption
that both parameters are universal.

\section{Acknowledments}

The author would like to thank my ZEUS collaborators for their efforts to
produce the physics results presented at the conference, the ZEUS
management for giving her an opportunity to report them here and the
organizers for their hospitality.


\begin{thebibliography}{99}
\bibitem{1} ZEUS Coll., U Holm (ed.), {\it{The Zeus Detector}}. Status report (unpublished)\\
            DESY(1993), available on http://www-zeus.desy.de/bluebook/bluebook.html.
\bibitem{2} H1 Collaboration, C. Adloff {\it{et al.}}, \emph{Nucl. Phys.}  {\bf B 504} (1997) 3;\\
            ZEUS Collaboration, M. Derrick {\it{et al.}}, \emph{Z. Phys.} {\bf C 67} (1995) 93.
\bibitem{3} ZEUS Collaboration, S. Chekanov {\it{et al.}}, \emph{JHEP} {\bf 06} (2008) 061.
\bibitem{4} Z. Koba, H. B. Nielsen and P. Olesen, \emph{Nucl. Phys.}  {\bf B 40} (1972) 317.
\bibitem{5} TASSO Coll., W. Braunschweig {\it{et al.}}, \emph{Z. Phys.} {\bf C 45} (1989) 193.
\bibitem{b6} DELPHI Coll., P. Abreu {\it{et al.}}, \emph{Z. Phys.} {\bf C 50} (1991) 185.
\bibitem{b7} OPAL Coll., P. D. Acton {\it{et al.}}, \emph{Z. Phys.} {\bf C 35} (1991) 539.
\bibitem{b8} see Chekanov {\it{et al.}}, \emph{JHEP} {\bf 06} (2008) 061 and ref. [37] ibidem.
\bibitem{b9} ZEUS Collaboration, S. Chekanov {\it{et al.}}, \emph{JHEP} {\bf 08} (2009) 077.
\bibitem{b10} see S. Chekanov {\it{et al.}}, \emph{JHEP} {\bf 08} (2009) 077 and ref. [1] ibidem.
\bibitem{b11} Y. I. Azimov {\it{et al.}}, \emph{Z. Phys.} {\bf C 27} (1985) 65.
\bibitem{b12} see S. Chekanov {\it{et al.}}, \emph{JHEP} {\bf 08} (2009) 077 and refs. [3-9] ibidem.
\bibitem{b13} ZEUS Collaboration, J. Brietweg {\it{et al.}}, \emph{Eur. Phys. J.} {\bf C 6} (1999) 43.
 \bibitem{b14} S. Catani {\it{et al.}}, \emph{Nucl. Phys.}  {\bf B 406} (1993) 187.
\end{thebibliography}
\end{document}